\documentclass[epsfig,12pt,onecolumn]{article}
\usepackage{amsfonts}
\usepackage{amssymb}
\usepackage{amsmath}
\usepackage{multicol}
\usepackage{graphicx}
\usepackage{float}
\usepackage{caption}
\usepackage[top=2cm,left=2cm,right=2cm,bottom=2.5cm,includeheadfoot]{geometry}

\input{tcilatex}
\input{tcilatex}
\makeatletter \@addtoreset{equation}{section}

\def \be{\begin{equation}}
\def \ee{\end{equation}}
\def \bea{\begin{eqnarray}}
\def \eea{\end{eqnarray}}

\newcommand{\nc}{\newcommand}
\nc{\al}{\alpha} \nc{\bib}{\bibitem} \nc{\la}{\lambda}
\nc{\C}{\mbox{\hspace{1.24mm}\rule{0.2mm}{2.5mm}\hspace{-2.7mm} C}}
\nc{\R}{\mbox{\hspace{.04mm}\rule{0.2mm}{2.8mm}\hspace{-1.5mm} R}}

\begin{document}

\title{Particle-antiparticle in 4D charged Einstein-Gauss-Bonnet black hole}
\author{M. Bousder$^{1}$\thanks{%
mostafa.bousder@gmail.com}, M. Bennai$^{1,2}$\thanks{%
mohamed.bennai@univh2c.ma} \\
$^{1}${\small Lab of High Energy Physics, Modeling and Simulations,}\\
\ {\small Faculty of Science, Mohammed V University in Rabat, Morocco}\\
$^{2}${\small Quantum Physics and Applications Team, LPMC, Faculty of
Science Ben M'sik,}\\
{\small Casablanca Hassan II University, Morocco}}
\maketitle

\begin{abstract}
We study the charge of the 4D-Einstein-Gauss-Bonnet black hole by a negative
charge and a positive charge of a particle-antiparticle pair on the horizons 
$r_{-}$ and $r_{+}$, respectively. We show that there are two types of the
Schwarzschild black hole. We show also that the Einstein-Gauss-Bonnet black
hole charge has quantified values. We obtain the Hawking-Bekenstein formula
with two logarithmic corrections, the second correction depends on the
cosmological constant and the black hole charge. Finally, we study the
thermodynamics of the EGB-AdS black hole.

\textbf{Keywords:} Black hole, Entropy, Einstein-Gauss-Bonnet gravity.
\end{abstract}

\section{Introduction}

It's well known that the Gauss-Bonnet gravity is introduced only in case $%
D\succ 4$ or more. In four-dimensional spacetime, the GB term does not make
contributions to the gravitational dynamics, which makes the 4-dimensional
minimally coupled GB gravity is hard to obtain. Recently, D. Glavan and C.
Lin \cite{G1} proposed a novel 4-dimensional Einstein-Gauss-Bonnet (EGB)
gravity, which has attracted great attention. An intriguing idea of D.
Glavan and C. Lin is to multiply the GB term by the factor $1/(D-4)$ before
taking the limit. Offers a new 4-dimensional gravitational theory with only
two dynamical degrees of freedom by taking the $D\longrightarrow 4$ limit of
the Einstein-Gauss-Bonnet gravity in $D\succ 4$ dimensions \cite{G2}, which
is in contradiction with Lovelock theorem \cite{G3}. The four-dimensional
symmetrical static and spherical black hole solution in EGB gravity were
obtained \cite{G4}, also, solutions of a charged black hole \cite{G5}. There
has been lots of discussions about the self-consistency of the 4D EGB
gravity. It was shown in several papers that perhaps the $D\succ 4$\ limit
is not clearly defined, and several ideas have been proposed to remedy this
in-consistency \cite{G6,G7,G8}. Many researchers have studied particles in
the geometry of a black hole, precisely, the EGB black hole \cite{114,114A}.
Juan Maldacena and Leonard Susskind devised a theory linking two phenomena
both discovered by Einstein: \textquotedblleft Einstein-Rosen
bridges\textquotedblright\ (or wormholes) and quantum entanglement \cite{100}%
. According to them, if we move the two entangled particles apart would
amount to digging an ER bridge around a single particle which would manifest
its properties in several places in space-time. This theory sheds light on
the problem of the EPR paradox which highlights the non-locality of quantum
mechanics, which he opposes to the principle of the locality which is the
basis of the theory of relativity. However, this $ER=EPR$ correspondence is
only demonstrated in a very simplified universe model, where gravity is
generated in the absence of mass. The Hawking radiation of a black hole is a
scrambled cloud of radiation entangled with the black hole. The connection
between the laws of black hole mechanics with the corresponding laws of
ordinary thermodynamic systems has been one of the remarkable achievements
of theoretical physics \cite{100A}. In fact, the consideration of a black
hole as a thermodynamic system with a physical temperature and entropy
provides a deep insight to understand its microscopic structure. The study
of EGB gravity becomes very important because it provides a broader setup to
explore many conceptual questions related to gravity. This theory of gravity
similar to Einstein's gravity only benefits from the first and second order
derivatives of the metric function in field equations. there is also a great
connection between EGB gravity and the AdS / CFT demonstration
correspondence \cite{133}. \newline
Throughout the paper, we use the unit system where the speed of light $c=$
the gravitational constant $G_{N}$ $=$ \ the vacuum permittivity $4\pi
\varepsilon _{0}=1$.

\section{Charged Einstein-Gauss-Bonnet black hole}

Consider now the charged Einstein-Gauss-Bonnet theory in D-dimensions with a
negative cosmological constant \cite{115,116}

\begin{equation}
I=\frac{1}{16\pi }\int d^{D}x\left( R-2\Lambda +\frac{\alpha }{D-4}G-F_{\mu
\nu }F^{\mu \nu }\right)  \label{EG1}
\end{equation}%
where $\alpha $ is a finite non-vanishing dimensionless GaussBonnet coupling
have dimensions of $\left[ length\right] ^{2}$, that represent ultraviolet
(UV) corrections to Einstein theory, $F_{\mu \nu }=\partial _{\mu }A_{\nu
}-\partial _{\nu }A_{\mu }$ is the Maxwell tensor and $l$ is the AdS radius
and%
\begin{equation}
G=R^{2}-4R_{\mu \nu }R^{\mu \nu }+R_{\mu \nu \rho \sigma }R^{\mu \nu \rho
\sigma }  \label{EG2}
\end{equation}%
\begin{equation}
\Lambda =-\frac{\left( D-1\right) \left( D-2\right) }{2l^{2}}  \label{EG3}
\end{equation}%
by solving the field equation we obtain the black hole solution%
\begin{equation}
ds^{2}=-f(r)dt^{2}+\frac{1}{f(r)}dr^{2}+r^{2}\left( d\theta ^{2}+\sin
^{2}\theta d\phi ^{2}\right)  \label{EG4}
\end{equation}%
Taking the limit $D\longrightarrow 4$, we obtain the exact solution in
closed form%
\begin{equation}
-g_{00}=f(r)\approx 1+\frac{r^{2}}{2\alpha }\left( 1-\sqrt{1+4\alpha \left( 
\frac{2M}{r^{3}}-\frac{Q^{2}}{r^{4}}-\frac{1}{l^{2}}\right) }\right)
\label{EG5}
\end{equation}%
this last solution solution could be obtained directly from the derivation
done in \cite{116A}. In the large $r$ limit, for two branches of solutions.
In the limit of vanishing mass and charge \cite{116B}, for ($l^{-2}\sim 0$)\
we find\newline
\begin{equation}
-g_{00}=2\frac{r^{2}-2Mr+Q^{2}+\alpha }{r^{2}+2\alpha +\sqrt{r^{4}+4\alpha
\left( 2Mr-Q^{2}\right) }}  \label{EG6}
\end{equation}%
with $2M$ is the Schwarzschild radius. From this expression, we notice that,
in the limit $\alpha \longrightarrow 0$, the metric reduce to the
Reissner--Nordstr\"{o}m black hole solution. If $\alpha \prec 0$\ the
solution is still an AdS space, if $\alpha \succ 0$\ the solution is a de
Sitter space, \cite{116}. The event horizon in spacetime can be located by
solving the metric equation: $g_{00}=0$. To find the black hole horizon we
have to solve this equation\newline
\begin{equation}
r^{2}-2Mr+Q^{2}+\alpha =0  \label{EG7}
\end{equation}%
The solutions show that the event horizon is located at:%
\begin{equation}
r_{\pm }=M\pm \sqrt{M^{2}-Q^{2}-\alpha }  \label{EG8}
\end{equation}%
we notice that the solution behaves like the Reissner-Norstr\"{o}m (RN)
solution. The black hole event horizon is the largest root of the equation
above, $r_{+}$ is the black hole horizon \cite{G1}. However, the radius $%
r_{-}$ represents a horizon, which can be a horizon mirage or virtual
horizon. Therefore, to explain the presence of two horizon $r_{\pm }$, we
assume that we can represent the horizon $r_{-}$ as a reflection symmetric
of the horizon $r_{+}$ \cite{100}. There is only one case where the horizon $%
r_{+}$ corresponds to the horizon $r_{-}$, this case is equivalent to the
degenerate solution into a singularity when $\left\vert Q\right\vert =\sqrt{%
M^{2}-\alpha }$, which corresponding to an extremal black hole, and this
type of black hole which has a very low mass. We choose the same charge to
describe particle-antiparticle pair in the EGB black hole horizons. In this
paper, we propose that the two extremal entangled black holes are particles
and antiparticles which are placed just on the horizon of the EGB black
hole. The particle located near horizon $r_{+}$. On the other hand, the
antiparticle is the reflection of the particle on the horizon $r_{-}$. If
the particle and antiparticle are connected by a bridge (by an
Einstein-Rosen bridge), then the area of the bridge smaller than the area of
the black hole horizon. First, we define two charges of a
particle-antiparticle pair in the horizon

\begin{equation}
q_{\pm }=\pm \sqrt{M^{2}-\alpha }  \label{C8}
\end{equation}%
we will show later with a particular condition that the proposition of the
charges $q_{\pm }$ leads to an EGB entropy already obtained by other models.
Every particle on the horizon has a charge which depends directly on the
black hole mass; if the black hole mass larger, the charge of a horizon
particle will be more important. Each particle of the charge $q_{+}=+\sqrt{%
M^{2}-\alpha }$, is entangled with another antiparticle of the charge $%
q_{-}=-\sqrt{M^{2}-\alpha }$. The particle and antiparticle have opposite
electric charges $q_{+}$ and $q_{-}$, i.e. $CPT$ anticommutes with the
charges. If the number of particles $N$ on the horizon $r_{+}$ is limited,
the horizon charge will be $Q_{H}=0$. If we take that the number of horizon
particles $N\longrightarrow \infty $, the charge of the black hole is $%
Q_{H}\neq 0$. From Eqs.(\ref{EG8},\ref{C8})%
\begin{equation}
\left( r_{\pm }-M\right) ^{2}=q_{\pm }^{2}-Q^{2}  \label{C11}
\end{equation}%
according to this formula, the charges $q_{+}$ are located on the real
horizon, on the other hand, the charges of antiparticles $q_{-}$ are located
on a virtual horizon near the singularity. We also notice that $q_{\pm
}^{2}\succeq Q^{2}$. In the framework of the gravitational repulsion between
matter and antimatter, Eq.(\ref{C11}) behaves like virtual gravitational
dipoles \cite{118} in a black hole. If we assume that there is a complete
disappearance of an AdS black hole ($\alpha \prec 0$), we obtain the
position of the two charges%
\begin{equation}
r_{\pm }=q_{\pm }=\pm \sqrt{-\alpha }  \label{C13}
\end{equation}%
\newline
the problem of the negative radius $r_{-}$ indicates a disappearance of the
antiparticles with the disappearance of the black hole singularity, on the
other hand, the particles escape the singularity. This aspect is equivalent
to the position of the two horizons; the horizon $r_{-}$ exists on the
singularity and $r_{+}$ is the edge of the black hole. Which agrees with the
violation of CP symmetry between matter and antimatter \cite{119}. This
complete disappearance of the antiparticles looks like a scenario of the
disappearance of antimatter after the big bang. The relation (\ref{C13}) is
valid only for an AdS black hole, this may indicate that the disappearance
of the antiparticles is done on the second copy of AdS in other dimensions.
The electric potentials $\Phi _{+}$ and \ $\Phi _{-}$ arising from the
charge $q_{+}$and $q_{-}$ respectively, at a distance $r$ from the charge
given by 
\begin{equation}
\Phi _{+}(r)=\frac{q_{+}}{\left\vert r-r_{+}\right\vert }\text{ \ \ \ \ \ \ }%
\Phi _{-}(r)=\frac{q_{-}}{\left\vert r-r_{-}\right\vert }  \label{C14}
\end{equation}%
$\Phi _{+}$ and $\Phi _{-}$ are the conjugate (gauge independent) potentials
for the electric (and magnetic) $U(1)$ charges. Two opposite charges $q_{\pm
}$ have a potential of the electric dipole $\Phi _{+}+\Phi _{-}$ from%
\begin{equation}
\Phi _{+}(r)=\frac{\sqrt{M^{2}-\alpha }}{\left\vert r-M-\sqrt{%
M^{2}-Q^{2}-\alpha }\right\vert }\text{ \ \ \ \ \ \ }\Phi _{-}(r)=\frac{-%
\sqrt{M^{2}-\alpha }}{\left\vert r-M+\sqrt{M^{2}-Q^{2}-\alpha }\right\vert }
\label{C16}
\end{equation}%
In the extremal case $M^{2}=Q^{2}+\alpha $, we obtain $\Phi _{+}(r)=\Phi
_{-}(r)$. The two potential is canceled out for $M^{2}=\alpha $.

\section{Electric potential of EGB black hole}

Last papers proposed to study a charged particles near the Schwarzschild
black hole, or charged particle motion around magnetized Schwarzschild black
holes \cite{121,122,123,124}. We are also interested in studying the
particles (\ref{C8}) in the Schwarzschild horizon. We know that the
Schwarzschild metric is a solution for a black hole without electric charge
and angular momentum. In what follows we will use the previous results to
show that the Schwarzschild black hole is not charged ($Q=0$). \newline
we assume that the radius $r_{+}$ is equal to the Schwarzschild radius $2M$,
given by 
\begin{equation}
q_{\pm }=\pm M  \label{s1}
\end{equation}%
in this case, according to Eqs.(\ref{C11},\ref{C8}) one can obtain 
\begin{equation}
r_{-}=M\pm M  \label{s1a}
\end{equation}%
according to the two expressions (\ref{s1},\ref{s1a}), we can represent the
Schwarzschild black hole by two visions:\newline
If $r_{-}=0$, the Schwarzschild black hole consists of a negatively charged
singularity ($q_{-}=-M$) and a positively charged horizon ($q_{+}=M$),
therefore, if the number of particle and antiparticle is limited, the total
Schwarzschild black hole charge is zero, this case is similar to an
antimatter atom. If $r_{-}=2M$ (the horizon degenerates), the Schwarzschild
black hole consists of a neutral singularity and positive and negative
charges on the horizon ($q_{-}=-M$ ;$\ q_{+}=M$), therefore, for the
Schwarzschild horizon to be neutral; the number of positive charges must be
equal to the number of negative charges on the horizon. This result shows
that the Schwarzschild black hole, behaves like the neutral atom, but it
contains positive and negative charges. The Schwarzschild black hole
contains a negatively charged singularity $q_{-}$ and a positively charged
horizon $q_{+}$. We can express the electric potentials of negative and
positive charges in the Schwarzschild black hole by%
\begin{equation}
\Phi _{+}(r)=\frac{M}{\left\vert r\right\vert \left\vert 1-\frac{2M}{r}%
\right\vert }\text{ \ \ \ \ \ \ }\Phi _{-}(r)=-\frac{M}{\left\vert
r\right\vert }  \label{s2}
\end{equation}%
if we take $\alpha =0$ in Eq.(\ref{C16}), we also find the same potentials
above, what was mentioned by \cite{126}. Since the study of the charged
Einstein Gauss-Bonnet black hole shows a property similar to that of RN
black hole, like Eq.(\ref{EG8}), which shows that the parameter $\alpha $
creates a passage between the Schwarzschild black hole and RN black hole. We
can rewrite the Schwarzschild metric as%
\begin{equation}
ds^{2}=\frac{\Phi _{-}}{\Phi _{+}}dt^{2}-\frac{\Phi _{+}}{\Phi _{-}}%
dr^{2}+r^{2}\left( d\theta ^{2}+\sin ^{2}\theta d\phi ^{2}\right)  \label{s3}
\end{equation}%
the Schwarzschild metric depends on the potentials of particles and
antiparticles. In the Schwarzschild black hole, $N$ is limited and we write%
\begin{equation}
Q_{S}=q_{+}\sum_{n=0}^{N-1}\left( -1\right) ^{n}=0  \label{s4}
\end{equation}%
for these two relations above to be valid, the number $N$ must be even, i.e.
the number of particles equal to the number of antiparticles. But if $N$ is
odd, we get $Q=q_{+}$ or $Q=q_{-}$, this is at odds with an uncharged
Schwarzschild black hole. This shows that the number of particles on the
horizon is equal to the number of antiparticles in the Schwarzschild black
hole\textbf{\ }singularity, which corresponds exactly with the proposition (%
\ref{C8}) of entangled particles with antiparticles, one can't have on black
hole particles which are not entangled. Since the formula (\ref{EG8}) is a
generalization of the RN black hole horizons, this shows that the EGB black
hole charge is the generalization of charge of RN black hole (for $\alpha =0$%
) and zero charges of the Schwarzschild black hole. In the EGB black hole,
when $N$ $\longrightarrow \infty $, the analytic continuation of the Riemann
zeta function of $0$ ($\zeta (0)=1/2$) give%
\begin{equation}
Q_{EGB}=q_{\pm }\sum_{n=0}^{N-1}\left( -1\right) ^{n}=\pm \frac{1}{2}\sqrt{%
M^{2}-\alpha }  \label{s5}
\end{equation}%
\ \newline
this charge corresponds exactly with a physical event horizon because $%
Q_{EGB}$ check the condition of existence of the event horizon: $%
2Q_{EGB}\prec r_{s}$. We can also describe the charge of EGB black hole in
the case of $N\neq \infty $, with $N$ being odd, i.e. $Q_{EGB}=\pm q_{+}$.
Usually, we get

\begin{table}[]
\begin{center}
\begin{tabular}{|c|c|c|}
\hline
& $N\neq \infty $ & $N\longrightarrow \infty $ \\ \hline\hline
$Q_{EGB}^{-}$ & $-q_{+}$ & $-q_{+}/2$ \\ 
$Q_{S}$ & $0$ & $0$ \\ 
$Q_{EGB}^{+}$ & $q_{+}$ & $q_{+}/2$ \\ \hline
\end{tabular}%
\end{center}
\caption{The possible values of the EGB black hole charge as a function of
the charge of a particle.}
\end{table}

all values above of the black hole charge verify condition (\ref{C11}). In
the case where the charge $q_{+}=0$ is zero, the EGB black hole is
transformed into a Schwarzschild black hole. The values of the charge of $%
Q_{EGB}$ are quantified according to the constant values of $M$ and $\alpha $%
. We represent this quantification by the total charge of an Einstein
Gauss-Bonnet black hole is%
\begin{equation}
Q_{EGB}=m\sqrt{M^{2}-\alpha }\text{\ \ \ \ \ }m=\left\{
-1,-1/2,0,1/2,1\right\}  \label{s8}
\end{equation}%
Substitute Eq.(\ref{s8}) into Eq.(\ref{EG8}) and we obtain\newline
\begin{equation}
r_{\pm }=M\pm \sqrt{\left( 1-m^{2}\right) \left( M^{2}-\alpha \right) }
\label{s9}
\end{equation}%
the EGB black hole horizon is quantified according to the values of $m$. We
want to calculate the electric potential of the EGB black hole on the
horizon ($r=r_{+}$) from its charge (\ref{s8}) in the horizon $r_{+}$ (\ref%
{s9}), we obtain%
\begin{equation}
\Phi _{EGB}=\frac{Q_{EGB}}{r_{+}}=\frac{m\sqrt{M^{2}-\alpha }}{\sqrt{\left(
1-m^{2}\right) \left( M^{2}-\alpha \right) }+M}  \label{s13}
\end{equation}%
for the case where $m=0$, this potential becomes zero, which corresponds
exactly to the Schwarzschild black hole.

\section{EGB-AdS black hole thermodynamics}

\subsection{The black hole first law}

\bigskip We define the pressure \cite{127} of the cosmological constant (\ref%
{EG3}) for $D\longrightarrow 4$%
\begin{equation}
8\pi P=3l^{-2}\text{\ \ ;\ \ \ }\Lambda =-3l^{-2}  \label{D0}
\end{equation}%
We can express the ADM mass $M$ of the black hole in terms of $r_{\pm }$ by
solving Eq.(\ref{EG7}) for $r=r_{+}$ resulting in%
\begin{equation}
M=\frac{l^{-2}r_{+}^{4}+r_{+}^{2}+Q_{EGB}^{2}+\alpha }{2r_{+}}  \label{D1}
\end{equation}%
The Hawking temperature is easy to give by calculating surface gravity at
the horizon%
\begin{equation}
T^{2}=-\frac{1}{8\pi ^{2}}\nabla _{\mu }\xi _{\nu }\nabla ^{\mu }\xi ^{\nu }=%
\frac{1}{16\pi ^{2}}f^{\prime 2}(r_{+})  \label{D2}
\end{equation}%
where $\xi ^{\mu }$ is a killing vector, which for a static, spherically
symmetric case takes the form $\xi ^{\mu }=\partial _{t}^{\mu }$. $\xi _{\mu
}$ satisfies the Killing equation%
\begin{equation}
\nabla _{\mu }\xi _{\nu }+\nabla _{\nu }\xi _{\mu }=0
\end{equation}%
The Hawking temperature of the black hole can be calculated as

\begin{equation}
T=\frac{3l^{-2}r_{+}^{4}+r_{+}^{2}-Q_{EGB}^{2}-\alpha }{8\pi \alpha
r_{+}+4\pi r_{+}^{3}}  \label{D4}
\end{equation}%
if we suppose that the charges $q_{+}$ and $q_{-}$ are behave like a gas on
the black hole. The black hole first law reads \cite{127} 
\begin{equation}
dM=TdS+\Phi _{EGB}dQ_{EGB}+VdP+Ad\alpha  \label{D5}
\end{equation}%
\ \ \newline
The parameters $V$ and $A$ are the conjugate quantities of the pressure $P$
and GB coupling parameter $\alpha $, respectively.\ \newline
If we fix $P$ and $\alpha $ the Hawking-Bekenstein formula is then given by%
\begin{equation}
S=\int \frac{dM}{T}-\int \frac{\Phi _{EGB}}{T}dQ_{EGB}  \label{D6}
\end{equation}%
to describe the entropy according to the charges of the particles and
antiparticles present on the black hole, we have not fixed the charge $%
Q_{EGB}$ of the black hole. The solution of the first part of Eq.(\ref{D6})
already computed by \cite{116,127} as 
\begin{equation}
\int \frac{dM}{T}=\int_{0}^{r_{+}}\frac{1}{T}\left( \frac{\partial M}{%
\partial r^{\prime }}\right) dr^{\prime }=\pi r_{+}^{2}+2\pi \alpha \log
r_{+}^{2}\newline
+S_{0}  \label{D7}
\end{equation}%
\ \newline
where $S_{0}$ is an integration constant. Next, we use (\ref{s13}) to
calculate the second term in Eq.(\ref{D6}). We write the potential $\Phi
_{EGB}$ as a function of $Q_{EGB}$ for a fixed position $r=r_{+}$ we obtain 
\begin{equation}
\int \frac{\Phi _{EGB}}{T}dQ_{EGB}=\int \frac{Q_{EGB}}{Tr_{+}}dQ_{EGB}
\label{D12}
\end{equation}%
we substitute Eq.(\ref{D4}) into the last equation 
\begin{equation}
\int \frac{\Phi _{EGB}}{T}dQ_{EGB}=\int \frac{\left( 8\pi \alpha +4\pi
r_{+}^{2}\right) Q_{EGB}}{\left( 3l^{-2}r_{+}^{4}+r_{+}^{2}-\alpha \right)
-Q_{EGB}^{2}}dQ_{EGB}  \label{D13}
\end{equation}%
we consider that the expression $Q_{EGB}$ (\ref{s8}) is defined on a fixed
horizon (\ref{s9})%
\begin{equation}
\int \frac{\Phi _{EGB}}{T}dQ_{EGB}=-2\left( 2\pi \alpha +\pi
r_{+}^{2}\right) \log \left\vert Q_{EGB}^{2}-\left(
3l^{-2}r_{+}^{4}+r_{+}^{2}-\alpha \right) \right\vert +S_{1}  \label{D14}
\end{equation}%
with $S_{1}$ is an integration constant. Therefore, substitute Eqs.(\ref{D7},%
\ref{D14}) into Eqs.(\ref{D6}) we obtain%
\begin{equation}
S=\frac{A}{4}+2\pi \alpha \log \frac{A}{A_{0}}\ \newline
+\left( \frac{A}{2}+4\pi \alpha \right) \log \left\vert \frac{m^{2}}{A_{1}}%
\left( M^{2}-\alpha \right) -\frac{1}{4\pi A_{1}}\left( A-\Lambda A^{2}-4\pi
\alpha \right) \right\vert  \label{D15}
\end{equation}%
where $A_{0}$ and $A_{1}$\ are some constants with units of area, $A\equiv
4\pi r_{+}^{2}$ is the area of the event horizon of the black hole. This
equation generalizes the Hawking-Bekenstein (HB) formula by a two
supplementary logarithmic term \cite{128}, instead of just one additional
logarithmic form \cite{116,127}:%
\begin{equation}
S=\frac{A}{4}+2\pi \alpha \log \frac{A}{A_{0}}  \label{D16}
\end{equation}%
If we fix the EGB black hole charge we find the entropy according to the
model \cite{116}. We remark that the second entropy correction contains the
term $m$ which describes the states of the charge $Q_{EGB}$. We also remark
that there is the presence of the cosmological constant $\Lambda $ in the
second logarithmic term. Contrary to the model \cite{116}, the entropy above
described also by the cosmological constant with the charges $Q_{EGB}$, this
comes after varying $S$ with respect to $Q_{EGB}$. We calculated the
electric potential of EGB black hole, and by using this potential we write
the entropy with two logarithmic corrections. \newline

\subsection{Pressure, volume and temperature}

It is also possible to calculate the entropy (\ref{D15}) at finite
temperature $T=\beta ^{-1}$, by the use of expression (\ref{D4})\newline
which is also written as%
\begin{equation}
S=\frac{A}{4}+2\pi \alpha \log A\newline
+\frac{1}{2}\left( A+8\pi \alpha \right) \log \frac{B}{A_{1}}+S_{0}-S_{1}
\label{H01}
\end{equation}%
we define a new area%
\begin{equation}
B=\left\vert \beta ^{-1}r_{+}\left( A+8\pi \alpha \right) \right\vert
\label{H02}
\end{equation}%
we show that%
\begin{equation}
S=2\pi \alpha \log A\newline
e^{\frac{A}{8\pi \alpha }}B^{\frac{A+8\pi \alpha }{4\pi \alpha }}+S_{0}-S_{1}
\label{H03}
\end{equation}%
we study two different types of charged EGB-AdS black holes, allow for a
first order small-black-hole/large-black-hole (SBH/LBH), we choose for SBH a
surface of the form $A\ll 8\pi \alpha $, we choose for LBH a surface of the
form $A\gg 8\pi \alpha $, because the values of $\alpha $ included in the
interval $\left[ -1,1\right] $ \cite{133}. \newline
First, we assume that $A\ll 8\pi \alpha $, i.e. a space difference of AdS
space (since $\alpha \succ 0$). The entropy can be rewritten into the
following simple form%
\begin{equation}
S=2\pi \alpha \log A\newline
B^{2}+S_{0}-S_{1}\text{ \ and \ \ }B=8\pi \alpha \beta ^{-1}r_{+}
\label{H04}
\end{equation}%
we obtain a case where the subadditivity inequality is saturated for $%
S_{0}\preceq S_{1}$:%
\begin{equation}
S\equiv S(AB)\preceq S(A)+S(B)  \label{H05}
\end{equation}%
where 
\begin{equation}
S(A)=2\pi \alpha \log A\text{ \ and\ }S(B)=4\pi \alpha \log B  \label{H06}
\end{equation}%
Next, we assume that $A\gg 8\pi \alpha $, from Eq.(\ref{H03}) the entropy
can be rewritten into the following form%
\begin{equation}
S=\frac{A}{2}\log \frac{\sqrt{e}Ar_{+}}{\beta A_{1}}+2\pi \alpha \log \frac{A%
}{A_{0}}\newline
\label{H09}
\end{equation}%
According to $AdS/CFT$ correspondence, gravitational theories on $AdS_{2+1}$
space of radius $R$ are dual to $CFT_{2}$. \newline
for $r_{+}A=\beta A_{1}$, we obtain the same entropy in the EGB black hole
framework (\ref{D16}). \ \ \newline
Next, we assume that $A\sim 8\pi \alpha $%
\begin{equation}
S=A\log \frac{2e^{1/4}Ar_{+}}{\beta A_{1}}+\frac{A}{4}\log \frac{A}{A_{0}}%
\newline
\label{H10}
\end{equation}%
for $2r_{+}A=\beta A_{1}$, we obtain the same entropy in the EGB black hole
framework (\ref{D16}). According to the last two cases, we can transform the
two entropies (\ref{H09},\ref{H10}) into (\ref{D16}), which shows that there
is a possibility of return entropy with two logarithmic corrections (\ref%
{D15}) to entropy with a single correction. Therefore, we can find the
entropy (\ref{D16}) anyway by using the hypothesis (\ref{C8}) for the
condition of $A\succeq 8\pi \alpha $. This condition is also valid for an
AdS-EGB black hole: $\alpha \precsim 0$. Let us consider an asymptotically
AdS-EGB black hole spacetime ($\Lambda \prec 0$), the cosmological constant
corresponds to thermodynamic pressure with $8\pi P=-\Lambda $, and the
conjugate variable of $P$ corresponds to thermodynamic volume \cite{130} 
\begin{equation}
V=\left( \frac{\partial M}{\partial P}\right) _{S,Q,\alpha }=\frac{1}{3}%
r_{+}A
\end{equation}%
this conjugate variable was interpreted geometrically as an effective volume
for the region outside the EGB-AdS black hole horizon \cite{132}. For the
static black holes, the thermodynamic volume is only a function of the event
horizon. From (\ref{H09},\ref{H10}) we can introduce this condition 
\begin{equation}
A\succeq 8\pi \alpha \Longleftrightarrow \frac{1}{6}\beta A_{1}\preceq
V\preceq \frac{1}{3}\beta A_{1}  \label{H11}
\end{equation}%
the condition concerning the temperature is 
\begin{equation}
\frac{A_{1}}{3V}\preceq T\preceq \frac{A_{1}}{6V}  \label{H12}
\end{equation}%
the volume of AdS-EGB black hole checks the above relationship. Whether the
black hole surface checks $A\succeq 8\pi \alpha $, the maximum and minimum
values of the black hole volume depend on $\beta $. Therefore, the
temperature (\ref{D4}) is expressed as a function of the pressure (\ref{D0})
and $V$%
\begin{equation}
T=\frac{6PV}{A+8\pi \alpha }+\frac{r_{+}^{2}-Q_{EGB}^{2}-\alpha }{\left(
A+8\pi \alpha \right) r_{+}}  \label{H13}
\end{equation}%
we substitute Eqs.(\ref{s8},\ref{s9}) into the last equation, we get a
corresponding fluid equation of state: 
\begin{equation}
PV=\frac{A+8\pi \alpha }{6}T-\frac{1}{3}\sqrt{\left( 1-m^{2}\right) \left(
M^{2}-\alpha \right) }  \label{H14}
\end{equation}%
which leads to%
\begin{equation}
PV=\frac{A+8\pi \alpha }{6}T-\frac{1-m^{2}}{m^{2}}\frac{Q_{EGB}^{2}}{3\left(
r_{+}-M\right) }  \label{H15}
\end{equation}%
this equation is in good agreement with the equation $P=f(T,V)$ found by 
\cite{132} for a charged AdS black holes ($\alpha =0$). We use the specific
volume $\upsilon =2r_{+}l_{P}^{2}\equiv 2r_{+}=6V/N$, where $l_{P}=\sqrt{%
\hbar G/c^{3}}\equiv 1$ is the Planck length and $N=A/l_{P}^{2}$ is the
number of states associated with the horizon \cite{132}. 
\begin{equation}
P=\left( 1+\frac{8\pi \alpha }{A}\right) \frac{T}{\upsilon }-\frac{1-m^{2}}{%
m^{2}\left( 1-\frac{2M}{\upsilon }\right) }\frac{4Q_{EGB}^{2}}{\upsilon ^{2}N%
}  \label{H16}
\end{equation}%
Note that for $m=\pm 1$, we obtain the ideal gas law. We can show the Van
der Waals equation for $A\gg 8\pi \alpha $:%
\begin{equation}
\left( P+\frac{a}{\upsilon ^{2}}\right) \left( \upsilon -b\right) =T
\label{H17}
\end{equation}%
\begin{figure}[]
\centering
\includegraphics[width=11cm]{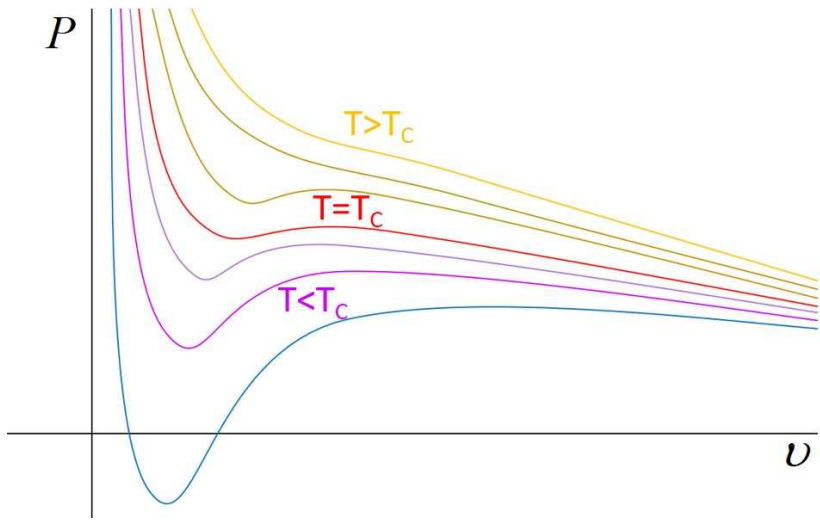}
\caption{Curves of some values $P-V$ isotherms of the van der Waals equation
of state.}
\label{graph}
\end{figure}
These curves are interpreted as follows:

for a temperature $T\succ T_{c}$ the fluid is only stable under one phase:
the supercritical fluid;

for a temperature $T\prec T_{c}$ the fluid is stable under a single-phase,
gas or liquid, or present simultaneously in two phases in equilibrium, gas
and liquid. the formula (\ref{H17}) yields%
\begin{equation}
a=\frac{4q_{+}r_{+}}{N}\sqrt{1-m^{2}}\succeq 0\text{ \ \ \ \ \ \ \ \ }\frac{b%
}{\upsilon }=\frac{8\pi \alpha }{A}  \label{H18}
\end{equation}%
where the parameter $a$ measures the attraction between particles and the
parameter $b$ corresponds to the volume of fluid of particles. We notice
that $b$ is less than $\upsilon $ because $A\gg 8\pi \alpha $. When $%
N\longrightarrow \infty $, the attraction between particles in the fluid
will be zero $a=0$. If there is a non-minimal coupling between
particle-antiparticle pair, the attraction will be zero for $%
N\longrightarrow \infty $. For an\textbf{\ }EGB-AdS black hole with $N\neq
\infty $ (table 1), the parameter $a$ will be zero.

\section{Conclusion}

In this paper, we have considered charged 4D EGB-AdS black holes as a
working substance. We studied the relationship between positive and negative
charges with the black hole in 4D EGB theory based on the work of Glavan \&
Lin \cite{G1}.\ We have assumed a particular shape of the charges present on
the black hole by the degenerate solution. We calculated the electric
potentials of particles and antiparticles for the EGB black hole, then we
deduce the potentials of these charges for the Schwarzschild black hole. We
have shown that the charge of a Schwarzschild black hole is zero. We have
adopted that the EBG black hole is a generalization of the Schwarzschild
black hole. We have found in this case a simplified charge of EGB black hole
which takes district values. By using this potential we write the HB formula
with two logarithmic corrections. The second correction depends on the
discrete values of the EBG black hole charge also depends on the
cosmological constant. For $A\succeq 8\pi \alpha $, we can obtain the HB
entropy with a single correction from that with two corrections. In this
case, we studied the AdS-EGB black hole thermodynamics, this study allows
obtaining the Van der Waals equation.

\end{document}